# Aspirated capacitor measurements of air conductivity and ion mobility spectra

K.L. Aplin

Space Science and Technology Department, Rutherford Appleton Laboratory, Chilton, Didcot, Oxon OX11 0QX UK

**Abstract**

Measurements of ions in atmospheric air are used to investigate atmospheric electricity and particulate pollution. Commonly studied ion parameters are (1) air conductivity, related to the total ion number concentration, and (2) the ion mobility spectrum, which varies with atmospheric composition. The physical principles of air ion instrumentation are long-established. A recent development is the computerised aspirated capacitor, which measures ions from (a) the current of charged particles at a sensing electrode, and (b) the rate of charge exchange with an electrode at a known initial potential, relaxing to a lower potential. As the voltage decays, only ions of higher and higher mobility are collected by the central electrode and contribute to the further decay of the voltage. This enables extension of the classical theory to calculate ion mobility spectra by inverting voltage decay time series. In indoor air, ion mobility spectra determined from both the novel voltage decay inversion, and an established voltage switching technique, were compared and shown to be of similar shape. Air conductivities calculated by integration were: $5.3 \pm 2.5$ $fSm^{-1}$ and $2.7 \pm 1.1$ $fSm^{-1}$ respectively, with conductivity determined to be 3 $fSm^{-1}$ by direct measurement at a constant voltage. Applications of the new Relaxation Potential Inversion Method (RPIM) include air ion mobility spectrum retrieval from historical data, and computation of ion mobility spectra in planetary atmospheres.

1.  Introduction

Atmospheric molecular cluster-ions are formed by natural radioactive isotopes and cosmic rays, and are central to the electrical properties of air. The original "Gerdien condenser"[1] aspirated capacitor is a widely used instrument for terrestrial atmospheric ion measurements. It consists of a cylindrical outer electrode containing a coaxial central electrode, with a fan to draw air between the electrodes. With an appropriate bias voltage applied across the electrodes, a current flows which is proportional to the air conductivity. (The "conductivity measurement régime" requires an adequate ventilation speed to be maintained for the bias voltage selected; the régime's existence can be verified by an approximately Ohmic response in measured current to a changing bias voltage). Early ion measurements inferred the air conductivity from the rate of voltage decay or relaxation across the electrodes, using a gold-leaf or fibre electrometer[2]. As electronics technology developed, this technique was augmented by direct measurements of the current. Sophisticated contemporary instruments under computer control combine the "current measurement" and "voltage decay" measurement modes for self-calibration[3,4].

Surface measurements with modern instrumentation suggest that, although generally comparable, conductivities from the two measurement modes (*i.e.* ion current and voltage decay) are not always completely consistent[3]. The study reported in [3] is believed to be the first direct comparison of the two modes, and motivated reconsideration of the theoretical assumptions underlying air conductivity measurement with an aspirated capacitor in the voltage decay mode. This paper describes a new technique, the Relaxation Potential Inversion Method (RPIM), enabling ion mobility spectra to be retrieved from voltage decay measurements. Improved ion measurements are needed for solar-terrestrial physics, pollution studies and assessment of long-term geophysical changes in atmospheric electrical parameters[5].

2. **Classical theory of air conductivity measurement with an aspirated cylindrical capacitor**

The electrical conductivity of air $\sigma$ is the product of air ion number concentration $n$, and ion mobility $\mu$. The ion mobility spectrum $n(\mu)$ describes the distribution of ion number concentration with mobility, and $\mu$ is inversely related to the size and molecular mass of the cluster. Molecular ions with $0.5 < \mu < 3$ cm$^2$V$^{-1}$s$^{-1}$ are conventionally defined as "small ions", as their size is limited by thermodynamic constraints on their lifetime, which generally inhibit ion growth to $\mu \sim 0.5$ cm$^2$V$^{-1}$s$^{-1}$ [6]. Applying these limits, the air conductivity due to positive or negative ions $\sigma_\pm$ is given by

$$\sigma_\pm = e \int_{0.5 cm^2 V^{-1} s^{-1}}^{3 cm^2 V^{-1} s^{-1}} \mu_\pm n_\pm(\mu) d\mu \qquad \text{Equation 1}$$

where $e$ is the charge on the electron, $\mu_\pm$ is the positive or negative ion mobility, and $n_\pm(\mu)$ the number of positive or negative ions with a given mobility. Because of the large differences between the mobility of small ions and aerosol particles, it is usually possible to assume that only small ions contribute to the conductivity, except in very polluted air. In this case, the lower limit of the integral can change, and intermediate and large ions of lower mobility may also be abundant enough to contribute[7]. Mohnen defined the mean mobility $\overline{\mu}$ as the mode of the ion distribution[8], usually 1.3-1.6 $cm^2V^{-1}s^{-1}$ for positive and 1.3-1.9 $cm^2V^{-1}s^{-1}$ for the chemically different negative ions[6]. At typical ionisation rates of 10 $cm^{-3}s^{-1}$, surface continental atmospheric ion concentrations are ~100-2000 $cm^{-3}$, and typical air conductivity can therefore vary considerably from ~2-100 $fSm^{-1}$. The determining influences on surface conductivity are aerosol pollution number concentrations (aerosol reduces the air conductivity, except in highly polluted air)[6,9] and the ion production rate from cosmic rays and geological sources.

The electrical mobility of small ions and aerosol particles differ by several orders of magnitude, so it is usually assumed that the principal contributions to the air conductivity are from ions with the same $\bar{\mu}$, and total number concentration $N$. Equation 1 is therefore commonly simplified to

$$\sigma = \sigma_+ + \sigma_- = n_+ e \bar{\mu}_+ + n_- e \bar{\mu}_- = Ne\bar{\mu}.$$

Equation 2

For an aspirated cylindrical capacitor operating in Current Measurement mode, with sufficient ventilation to ensure ions reaching the central electrode are constantly replenished, the ion current at the central electrode is proportional to the unipolar air conductivity. The motion of an ion in the radial electric field between the cylindrical capacitor's electrodes can be used with Equation 2 to derive the conductivity due to positive or negative ions $\sigma_\pm$ from the current $i$ at the central electrode arising from bias voltage $V_\pm$

$$\sigma_\pm = \frac{\varepsilon_0 i}{CV_\pm}.$$

Equation 3

The capacitance term $C$ accounts for radial electric field variations within the electrode system, usually found empirically to allow for connection and end effects[4], and $\varepsilon_0$ is the permittivity of free space. Full derivations of Equation 3 are given in [9], [10] and [11].

In the Voltage Decay mode, the voltage established across the capacitor electrodes decays due to the current *i* flowing through the air, of resistance *R*, Figure 1. If the instantaneous charge stored by the capacitor is *Q*, elementary circuit analysis gives

$$\frac{dQ}{dt}R = -\frac{Q}{C} = V,$$  Equation 4

As described in [11], Gauss's Law relates *i* to the air conductivity $\sigma$ by

$$i = -\frac{\sigma Q}{\varepsilon_0}.$$  Equation 5

Substituting Eq 5 into Eq 4, and differentiating with respect to time, gives

$$\frac{dV}{dt} = -\frac{\sigma}{\varepsilon_0}V.$$  Equation 6

If $\sigma$ is constant, the solution of Equation 6 gives the instantaneous voltage at a time *t*, *V(t)*, for an initial applied voltage $V_0$,

$$V(t) = V_0 \exp\left(\frac{-t}{\tau}\right),$$  Equation 7

where $\tau$ is a time constant = $\varepsilon_0/\sigma$, so Equation 7 can be rewritten as

$$V(t) = V_0 \exp\left(\frac{-t\sigma}{\varepsilon_0}\right),$$  Equation 8

Equation 8 has been the standard expression for calculating air conductivity from voltage decay measurements throughout the history of air ion instrumentation, using $\tau$ determined from a time series of voltage data[2,3,12].

It is important to emphasise that the Voltage Decay and Current Measurement modes are fundamentally different in the mobility of ions assumed to be selected. For both modes, the ion mobility contributing to the measurement is assumed to be constant, but the longer duration required for Voltage Decay measurements requires the assumption that the mobility spectrum being sampled is constant for longer than the measurement period. If very long decay timescales are considered, this implies that such measurements could be susceptible to temporal or other fluctuations in ion mobility.

3. **Modification to the classical theory of the Voltage Decay mode**

In calculations of the conductivity, the critical mobility $\mu_c$ is assumed to represent the minimum mobility of ion contributing to the measurement[10]. It is defined from consideration of the motion of ions in the radial electric field at a ventilated capacitor (Figure 1)[10]. For a cylindrical geometry, $\mu_c$ is a function of ventilation speed $u$, length $L$, central and outer electrode radii $a$ and $b$ and bias voltage $V$ given by:

$$\mu_c = \frac{ku}{V},  \quad \text{Equation 9}$$

[10] where $k$ is a geometrical constant:

$$k = \frac{(a^2 - b^2)\ln(a/b)}{2L}. \quad \text{Equation 10}$$

Critical mobility is a function of bias voltage, and therefore ion mobility spectra can be found by changing the voltage at the central electrode[13]. It is possible for some ions with mobility lower than the critical mobility to enter the cylindrical capacitor, but this effect is negligible except in polluted air with very high concentrations of larger charged particles[14]. In this paper it is assumed that only ions with mobility greater than the critical mobility contribute to the measurements. The implications of this assumption will be discussed in section 5.

During a Voltage Decay measurement, both the voltage, and therefore the critical mobility (Equation 9) vary continuously. As a consequence, the decaying voltage across the capacitor's electrodes changes the mobility distribution of the ions contributing to charge exchange. This modulation of the ion mobility spectrum, and therefore, from Equation 1, air conductivity, invalidates the assumption used in the derivation of Equation 7, that the ion spectrum selected for measurement remains constant. The behaviour of an aspirated capacitor in voltage decay mode cannot be completely described by Equation 7: in an ideal instrument, differences from Equation 7 will arise from changes in critical mobility during the decay.

Differences from the exponential decays predicted by Equation 7 appear detectable in measurements of voltage decays in atmospheric air, following a series of measurements made over several months which rarely showed the exponential decays expected based on classical assumptions[9]. Additionally, past voltage decays measured in the free troposphere were also non-exponential[12]. Figure 2 shows exponential fits to typical voltage decay time series in surface atmospheric air. Natural variability in the measurement is expected to cause some fluctuations in the time series, particularly in the polluted boundary layer, but the existence of free tropospheric non-exponential decays is more difficult to explain using the theory outlined in section 2. Classical theory can be modified to account for the variation in critical mobility during voltage decay measurements.

## 4. Computing ion spectra from voltage decay measurements

As conductivity is effectively the mobility integral over the ion spectrum (Equation 1), every voltage decay time series contains, in principle, ion spectrum information. The relationship between voltage decay measurements and the ion spectrum information can be determined by substituting Equation 1 for the conductivity term in Equation 6, and evaluating the resulting integral in two parts. The first part is with respect to mobility, from the maximum ion mobility in the air $\mu_m$ down to the critical mobility evaluated at time $t$, $\mu_c(t)$. The second part is evaluated from $\mu_c(t)$ to the critical mobility at the start of the decay $\mu_c(0)$, written as

$$\ln\frac{V_t}{V_0} = -\frac{e}{\varepsilon_0}\left[t\int_{\mu_c(t)}^{\mu_m}\mu n(\mu)d\mu + \int_{\mu_c(0)}^{\mu_c(t)}\mu n(\mu)t_c(\mu)d\mu\right].$$  Equation 11

Although Equation 11 is not generally analytically soluble, $n(\mu)$ can be found by using a finite difference numerical method, applicable for small changes. This permits calculation of ion spectra from voltage time series, which is the basis of the Relaxation Probe Inversion Method (RPIM). It can also be used for prediction of voltage decays from a given ion spectrum[9].

### 4.1. Numerical solution procedure

Writing the two integral terms in Equation 11 as $I$ and $M$ gives

$$-\frac{\varepsilon_0}{e}\ln\frac{V_t}{V_0} = tI + M.$$  Equation 12

During a finite, small time difference between $t_{j-1}$ and $t_{j+1}$ the voltage will have decayed slightly, and caused a small increase in the instrument's critical mobility. Thus Equation 12 can be written in finite difference form as

$$-\frac{\varepsilon_0}{e}\ln\frac{V_{j+1}}{V_{j-1}} = j\Delta I_j + \Delta M_j$$  Equation 13

with the changes in $I$ and $M$, $\Delta I_j$ and $\Delta M_j$ approximated at each time $j$ by:

$$\Delta I_j = [\mu_c(t_{j+1}) - \mu_c(t_{j-1})]n_j$$  Equation 14

$$\Delta M_j = \left\{ \left( \mu_c(t_{j+1}) - \mu_c(t_{j-1}) \right) \left( -\tau_j \ln\left( \frac{\mu_c(t_{j-1})}{\mu_c(t_{j+1})} \right) - t_j \right) \left[ \frac{1}{2} \left( \mu_c(t_{j-1}) + \mu_c(t_{j+1}) \right) \right] n_j \right. \quad \text{Equation 15}$$

where $\mu_c(t_j)$ is the critical mobility evaluated from the voltage (Equation 9) at each time $j$, and $\tau_j$ the instantaneous decay time constant. The incremental changes, $\Delta I_j$ and $\Delta M_j$ can be evaluated for each mobility strip of width $[\mu_c(t_{j+1}) - \mu_c(t_{j-1})]$, with mobility calculated using Equation 9 from $V(t_{j+1})$ and $V(t_{j-1})$. The ion concentration in each mobility strip $n_j$ can then be estimated. The inversion yields a mobility spectrum with $N$-1 points if the original voltage decay has $N$ points; the highest voltage (corresponding to the lowest ion mobility) does not have a corresponding spectral point because it provides the initial voltage $V_0$. The steps involved in the inversion procedure are summarised in Figure 3.

*4.2. Numerical example with a synthetic ion spectrum*

A numerical example demonstrates the inversion of a voltage decay generated from a synthetic ion mobility spectrum, broadly similar to the observed mean small ion spectrum[6]. A voltage decay time series for the aspirated cylindrical capacitor described in [3] and [9] was generated from the ion spectrum, Figure 4. Inversion of this voltage decay using the RPIM algorithm (Section 4.1) results in an identical spectrum to the original, indicating that the mathematical inversion is exact. However the typical shape of small ion spectrum is generated as an average over many measurements; the spectrum inverted numerically here was an average of 8615 hourly averaged ion spectra taken over 14 months[6]. The example demonstrates that the RPIM can correctly determine ion spectra from voltage decay data, but independent, and ideally synchronous, measurements of voltage decays and ion mobility spectra are necessary in practical evaluations. It should also be noted that the second term in Equation 11 is small compared to typical experimental uncertainty, although it has been included in this numerical example for exactness.

**5. Comparison of ion spectra measured in laboratory air**

The RPIM was verified by comparing ion mobility spectra calculated from voltage decays, to spectra computed from the established technique of voltage switching to vary the critical mobility[13,15]. The experiments were carried out in ambient indoor air, in a demonstrably stable atmospheric electrical environment (described in detail in [16]), using the computer-controlled Programmable Ion Mobility Spectrometer (PIMS), with sensing electrodes 0.25m long and of radii 11mm and 2mm, ventilated at 2.1 ms$^{-1}$ [3,4].

Voltage decay measurements were carried out using a Keithley 236 SourceMeter instrument to supply, and then measure (to a specified accuracy of to ± 0.025 % + 10mV), the voltage across the ventilated PIMS electrodes. Data was logged by a PC at 1Hz via a GPIB/IEEE interface. The maximum voltage supplied by the Keithley 236 is 110 V, corresponding to a critical mobility of 0.16 cm$^2$V$^{-1}$s$^{-1}$ ± 0.025 %. Three voltage decay events from nominally 110 V – 3 V, each of duration 2-3 hours were measured on 23 and 24 March 2005, Figure 5a.

The Current Mode ion spectrum measurements took place on 24-25 March 2005, at the same temperature and relative humidity as the voltage decay measurements. Voltage switching through a predetermined sequence of error checking modes (as in [4]), and 15 bipolar voltages from -20.8 V - 21.1 V was implemented in software, and data logged via the RS232 port. Equation 9 can be used to calculate critical mobilities; only results from the positive voltage sequence (3.9V - 21.1V) are used here, corresponding to positive ion critical mobilities in the range 0.9 - 3.5 cm$^2$V$^{-1}$s$^{-1}$. Errors in the critical mobility result principally from the calibration of the digital to analogue converter used to generate the bias voltage, ± ~10%. Ion currents were sampled at 1Hz and averaged over 20s for each voltage step, and 90s of empirically determined recovery time was allowed between each change in bias voltage. The mean currents are plotted as a function of bias voltage in Figure 5b. Combining Equation 2 and Equation 3 gives

$$n_\pm \approx \frac{\varepsilon_0 i}{CV_\pm e \overline{\mu}}, \qquad \text{Equation 16}$$

where $\overline{\mu}$ is the average mobility. Increasing the bias voltage in steps from $V_{i-1}$ (through $V_i$) to $V_{i+1}$ causes a change in critical mobility from $\mu_{c(i-1)}$ to $\mu_{c(i+1)}$, which will increase the magnitude of the ion current from $i_{(i-1)}$ to $i_{(i+1)}$. The ion concentration in the mobility range centred on $\mu_{ci}$, $n_i(\mu_{ci})$, can be written as

$$n_i(\mu_{ci}) = \frac{\varepsilon_0}{Ce|V_i|} \left| \frac{(i_{(i-1)} - i_{(i+1)})}{(\mu_{(i-1)} - \mu_{(i+1)})} \right|. \qquad \text{Equation 17}$$

Experimentally, the rate of change of current with critical mobility can be determined by using linear regression between a set of measured current and critical mobility values (calculated for each bias voltage using Equation 9). Methods for measuring the capacitance term in Equation 17 are described in [4] and [17]).

Both the RPIM and the voltage switching technique assume that no ions with mobility lower than the critical mobility can enter the cylindrical capacitor. However, some of these larger charged particles, which constitute the particulate space charge, can contribute to the measurement by drifting into the instrument and colliding with the sensing electrodes. The magnitude of the error from particulate space charge can be estimated by measuring the current at the central electrode with zero bias voltage applied, i.e. the current arising from particles unaffected by the electric field. This has been referred to previously as a "dynamic zero".[10]. A more rigorous approach to find the dynamic zero is to calculate the intercept of the bipolar current-voltage plots obtained during the voltage switching measurements, which is a time-averaged dynamic zero. This dynamic zero was subtracted to produce the *i-V* curve shown in Figure 5b, but corresponded to 2.7 pCm$^{-3}$ of negative space charge, or a maximum of 17 (singly charged) particles cm$^{-3}$.

The mean ion spectrum was calculated using the RPIM as in Section 4.1, and ion spectra from the voltage decay and current mode methods are shown in Figure 6. The spectra are similar in shape, and the mobility of the common peak is consistent with positive ion properties in the literature[8,11]. If the positive air conductivity due to small ions is calculated by integration across the mobility spectrum (Equation 1), then $\sigma_{\text{(current mode)}}$ = 2.7 ± 1.1 fSm$^{-1}$, with the error determined from the variability in current measurements. Mean $\sigma_{\text{(voltage decay mode)}}$ = 5.3 ± 2.5 fSm$^{-1}$, where the error is the standard deviation across the three measured spectra. This is the same as positive air conductivity measured at the same location by an aspirated cylindrical capacitor operating in current mode at a constant bias voltage, ~3 fSm$^{-1}$ [16]. The co-located peaks, well-correlated spectral shape and consistent integral spectra all give confidence in the RPIM approach, although there is some disagreement between the concentrations calculated at the extremes of the small ion range, $1.5 > \mu > 2.3$ cm$^2$V$^{-1}$s$^{-1}$, The comparison of the new RPIM spectra with spectra obtained using the well-established technique shows firstly that the inversion generates reproducible and realistic ion mobility spectra, and secondly that the air conductivity computed by integration across the spectrum is comparable with that found in the same environment by a different method.

## 6. Discussion

The classical theory of air conductivity measurement for voltage decays from an aspirated capacitor can be modified to correct for the assumption that the critical mobility of ions sampled by the instrument does not vary during the decay. The spectral information extracted from voltage decay data with the RPIM can be used to calculate conductivity directly by integration rather than with erroneous simplifying assumptions, such as exponential decay.

RPIM has practical as well as theoretical advantages. Measurements of atmospheric ion spectra (reviewed in detail in [9]) are often obtained by varying the bias voltage to change the ion mobilities selected *e.g.* [13],[14]. As in the example described in Section 5, this requires dedicated electronics to switch the bias voltage, and sensitive current sensing. Both time and mobility resolution can be poor, as compromises must be made between the voltage size step, and the time the instrument rests at each bias voltage to obtain an averaged current. Adequate time must also be allowed for the circuitry to recover from switching transients. RPIM exploits continuous voltage decay, which avoids delays and errors from switching transients. Smooth voltage decays also minimise the sharp changes in electric field at the cylindrical capacitor inlet caused by bias voltage switching, which may perturb ion ingress and cause transient saturation in the instrument, briefly preventing any measurements[17]. Voltage decay measurements need only the simplest single-channel logging equipment and are ideally suited to remote in situ sensing applications, such as balloon-borne measurements[18]. Data processing to compute the ion spectrum and integrated conductivity would typically be carried out off-line.

The RPIM assumes that no intermediate or large ions can enter the cylindrical capacitor; however, it can be seen from Section 5 that the magnitude of uncertainty introduced by this assumption is much lower than the variability between individual spectra. RPIM can also be used for extraction of spectral information from historical atmospheric electrical data sets. An important new application is the inversion of voltage decay measurements made in the atmospheres of other planets, such as from the European Space Agency Huygens probe which used voltage relaxation techniques during the first *in situ* measurements of Titan's atmosphere[19].

**Acknowledgements**

I thank Dr C.F. Clement for his assistance with mathematics, and Dr R.G. Harrison for helpful discussions. J.G. Firth provided technical support, and the UK Natural Environment Research Council (NERC)/RAL Molecular Spectroscopy Facility (funded by a NERC New Investigators' grant NER/M/S/2003/00062) was used for some of the experimental measurements.

**Figure Captions**

Figure 1 Schematic of an aspirated cylindrical capacitor showing a plan view of the end of the tube (centre), a section through the tube (left) and the equivalent circuit (right). The motion of a charged particle through the tube is indicated (left). A charging voltage $V_0$ is applied and released to measure ions in the Voltage Decay mode (right).

Figure 2 Consecutive voltage decay time series measured in urban atmospheric air on June 12 1998. (The experimental apparatus and other results are described in detail in [3]). Coefficient of determination ($R^2$) values are shown to indicate the fraction of variance in a data set explained by an exponential model. For 39 voltage decays measured over two weeks, the mean $R^2$ was 0.9 with a standard deviation of ± 0.1 (the range of values was 0.31-1.00).

Figure 3  Flow chart illustrating the algorithm developed to invert voltage decay data to obtain an ion mobility spectrum.

Figure 4 Inversion of an artificial ion mobility spectrum, chosen to closely represent an atmospheric ion mobility spectrum. The experimental voltage decay time series the artificial spectrum would have generated is shown (inset), with a best fit line to an exponential. The voltage decay time series was then inverted back to an ion spectrum, and is plotted on the same axes as the original artificial spectrum.

Figure 5 a) Average of three voltage decays in indoor air measured with the Programmable Ion Mobility Spectrometer on 23-24 March 2005. b) Currents measured over a range of bias voltages at the same location on 24-25 March 2005. The y-axis error bars are the standard error of the mean.

Figure 6 Comparison of average indoor small positive ion spectra generated from the Relaxation Probe Inversion Method and by the established bias voltage switching technique. Typical errors in the x-axis mobility values are ±10%. Estimated errors in ion concentrations are ±30% for the switched voltage spectrum, and ±40% for the RPIM spectrum.

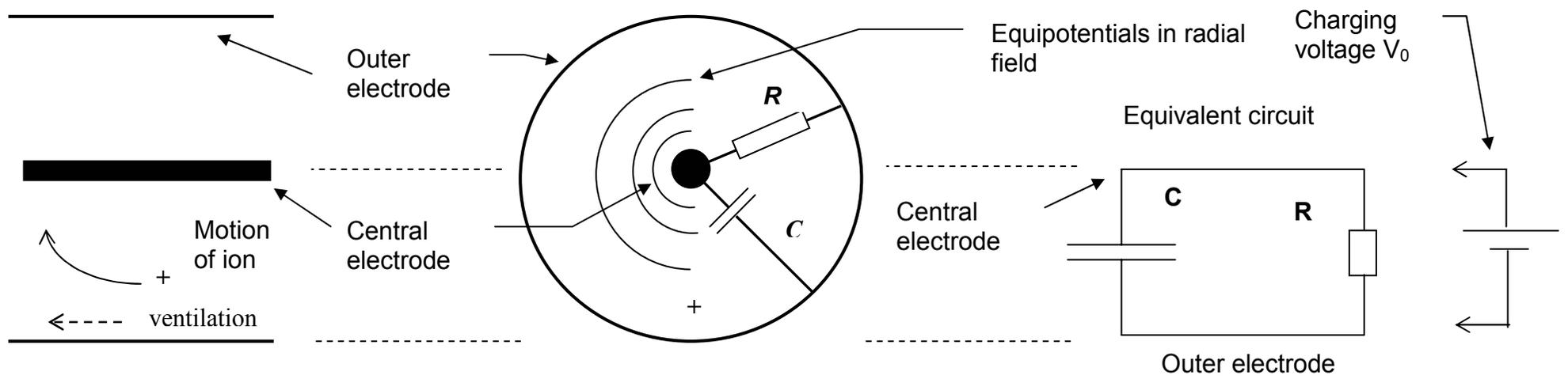

Figure 1

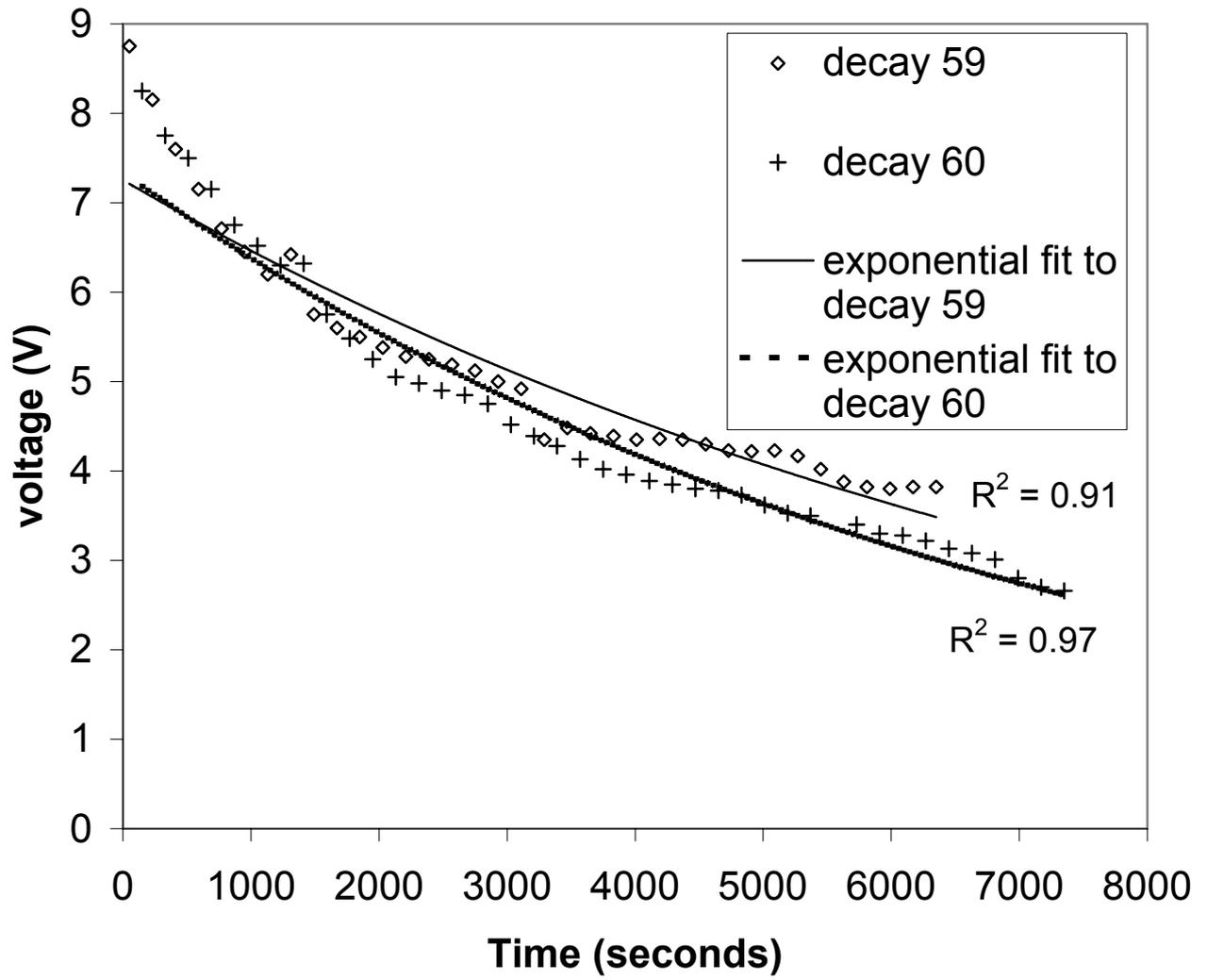

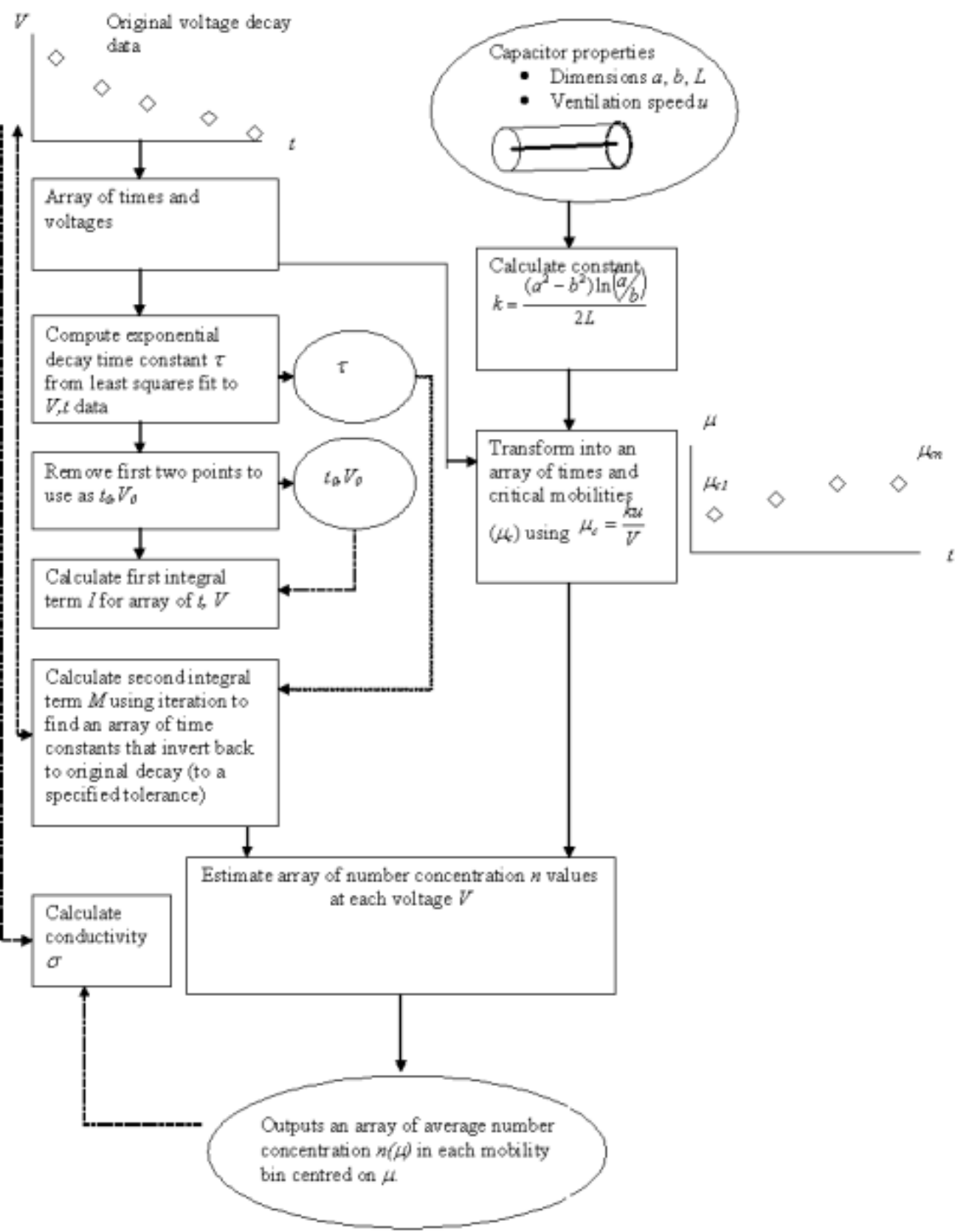

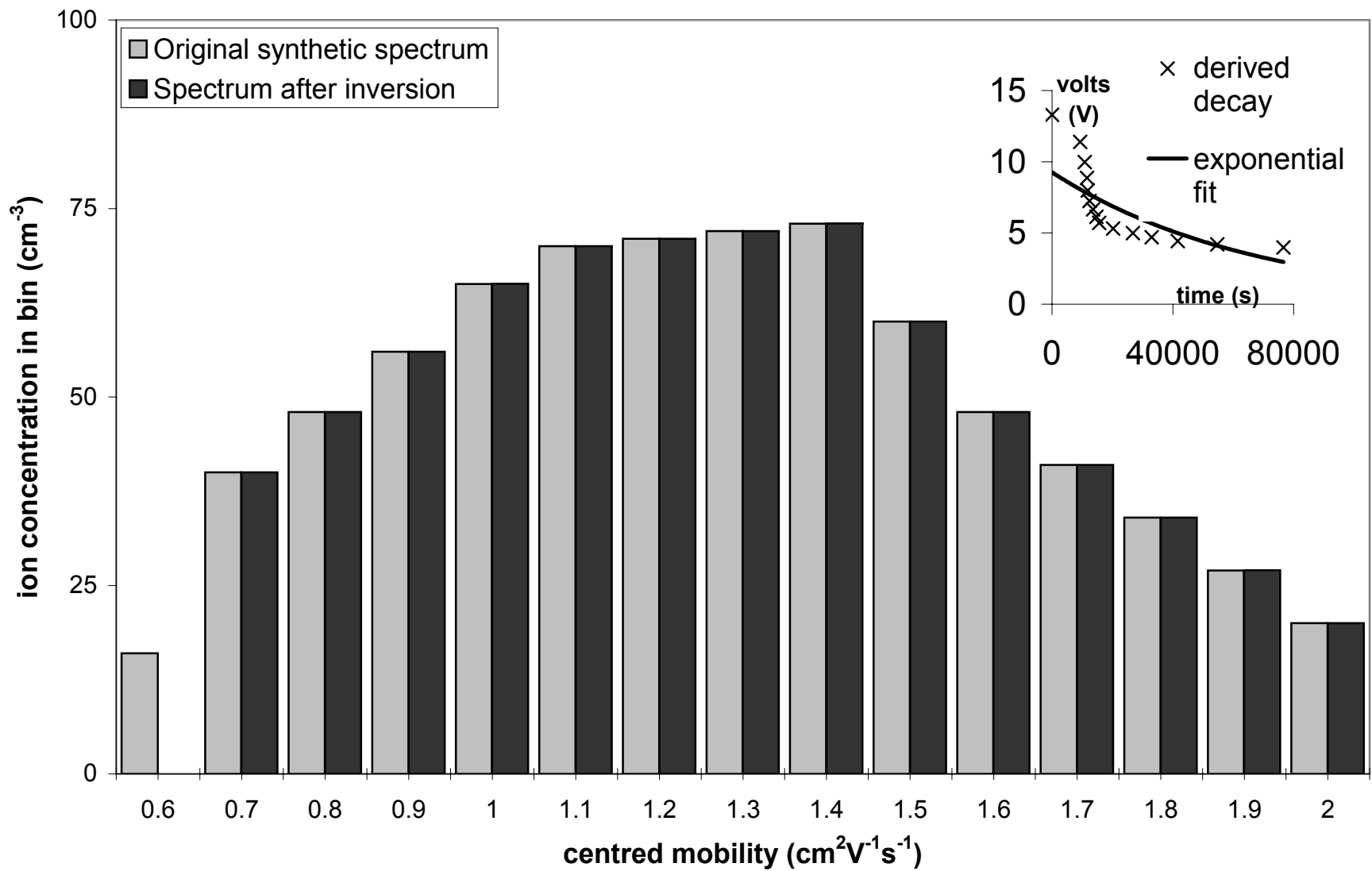

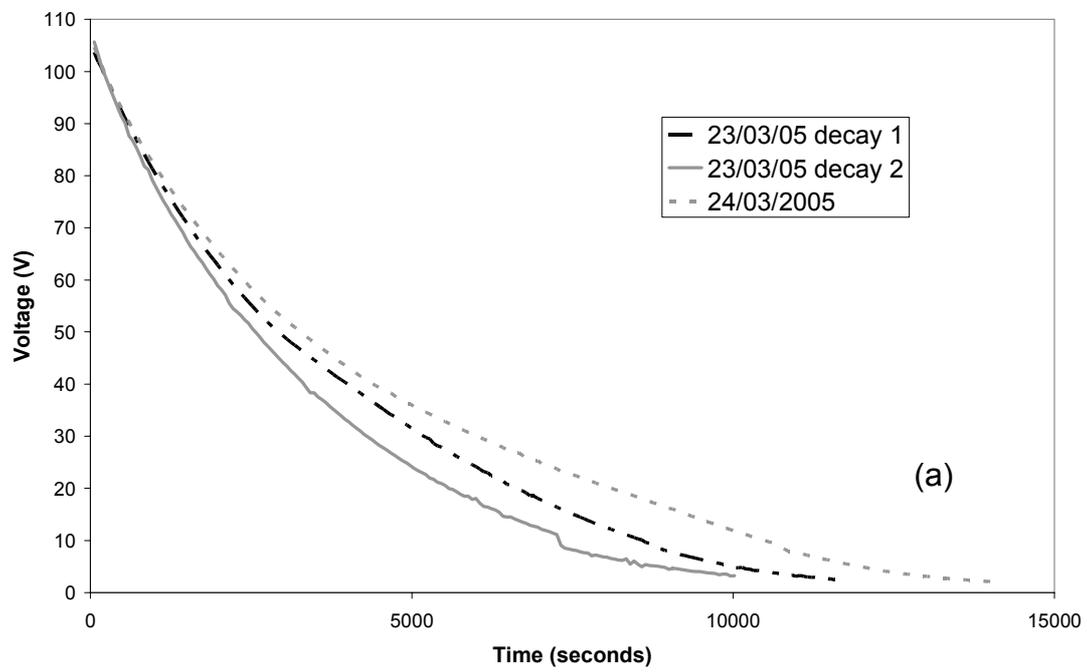

(a)

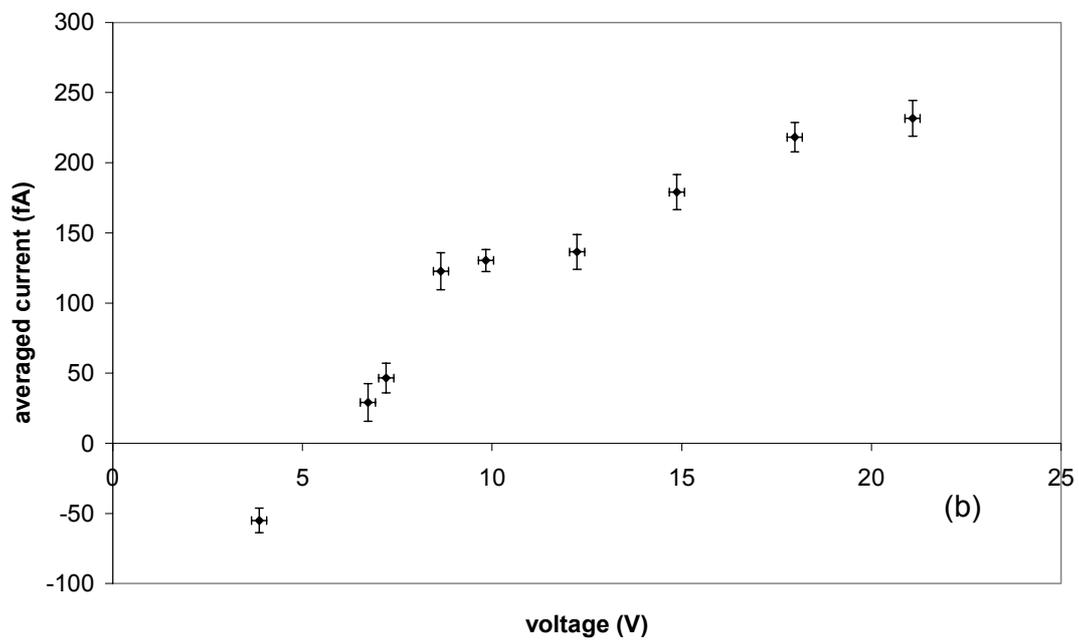

(b)

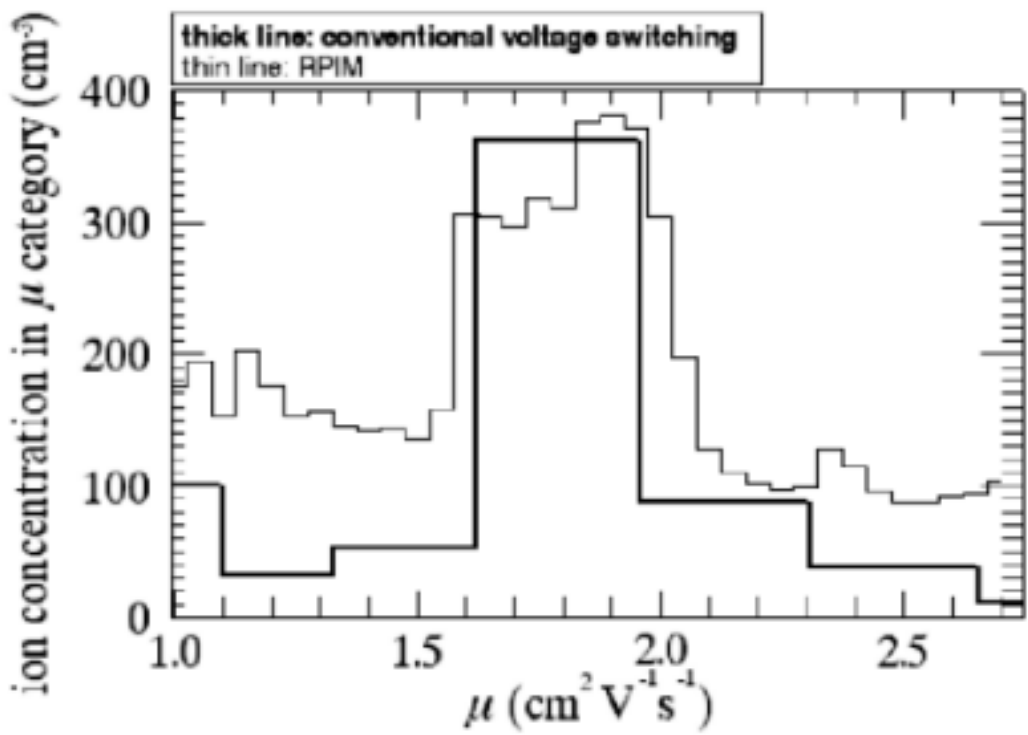